\documentclass[12pt]{iopart}
\usepackage{graphicx}
\usepackage{amssymb}
\usepackage{cite}
\usepackage{dcolumn}    
\usepackage{bm}         

\begin{document}

\title{       Jahn-Teller distortions and the magnetic order \\
              in the perovskite manganites }

\author{      Krzysztof Ro\'sciszewski$^1$ and
              Andrzej M. Ole\'s$^{1,2}$  }

\address{$^1$ Marian Smoluchowski Institute of Physics, Jagellonian
              University, \\ Reymonta 4, PL-30059 Krak\'ow, Poland }
\address{$^2$ Max-Planck-Institut f\"ur Festk\"orperforschung, \\
              Heisenbergstrasse 1, D-70569 Stuttgart, Germany }

\ead{roscis@th.if.uj.edu.pl; a.m.oles@fkf.mpg.de}

\date{\today}

\begin{abstract}
We introduce an effective model for $e_g$ electrons to describe
three-dimensional perovskite (La$_{1-x}$Sr$_{x}$MnO$_3$ and
La$_{1-x}$Ca$_{x}$MnO$_3$) manganites and study the magnetic and
orbital order on $4\times 4\times 4$ cluster using correlated wave
functions.
The model includes the kinetic energy, and on-site Coulomb
interactions for $e_g$ electrons, antiferromagnetic superexchange
interaction between $S=3/2$ core spins, and the coupling between
$e_g$ electrons and Jahn-Teller modes.
The model reproduces the experimentally observed magnetic order:
($i$) $A$-type antiferromagnetic phase in the undoped insulator
LaMnO$_3$, with alternating 
$e_g$ orbitals and with small Jahn-Teller distortions, changing to
a conducting phase at 32 GPa pressure, and ($ii$) ferromagnetic
order in one-eight doped La$_{7/8}$Sr$_{1/8}$MnO$_3$ and in quarter
doped La$_{3/4}$Sr$_{1/4}$MnO$_3$ compounds.
For half-doped La$_{1/2}$Ca$_{1/2}$MnO$_3$ one finds a competition
between a ferromagnetic conductor and the CE insulating phase;
the latter is stabilized by the Jahn-Teller coupling being twice
larger than for the strontium-doped compound.
Altogether, there is a subtle balance between all Hamiltonian
parameters and the phase diagram is quite sensitive to the precise
values they take. \\

\end{abstract}

\vskip 1cm

{\it Published in: J. Phys.: Condensed Matter {\bf 22}, 425601 (2010).}

\maketitle

\section{Introduction}

Much effort was put into understanding what is going on in doped
perovskite manganise oxides both on theoretical
\cite{dag01,millis,saitoh1,mizokawa,Ben99,key,popovic,hotta,kovaleva,ebata,daghofer,yin,lin08}
and experimental
\cite{morimoto,110K,sternlieb,park,maezono,jung,perring01,wilkins,wilkins06,chatterji,huang,senff,jprim} side.
Early expectations that simple physical mechanisms, such as involving
for example only Jahn-Teller (JT) interactions and Hund's exchange
but without Coulomb interaction, or pure electronic Coulomb
interactions while neglecting Hund's exchange, could explain the phase
diagram and the phase transitions observed situations in the manganites
were not confirmed \cite{dag01,hotta}. Also relatively early it was
realized that the core $t_{2g}$ electrons, even if dynamically passive,
play an essential role for the stability of magnetic and orbital order
\cite{daghofer,mizokawa}.
Altogether, the evidence has accumulated that there is a delicate
balance between several competing physical mechanisms which lead to
rather complex phase diagram of the perovskite manganites
\cite{dag01,daghofer,lin08}.

Investigation of doped manganites is challenging as the Jahn-Teller
distortions may lead to the charge order which will also favour
particular orbital order \cite{dag01,Ole10}.
To make the complex situation in doped manganites tractable in the
theory, several theoretical papers were focused
in past on single-layer and double-layer manganites as the description
of quasi two-dimensional (2D) systems was simpler than the one
necessary for three-dimensional (3D) doped perovskite manganites
(see for example \cite{daghofer} and the references therein).
Realistic 3D systems were more difficult to tackle at least on a better
basis than simple mean-field type approaches. Due to strong local
correlations, however, such approaches were not quite appropriate
to provide a better insight into the competition between different
possible physical mechanisms present in these strongly correlated
electron systems. The present paper is one attempt (among the others)
to fill up this gap.

Similar to model studies \cite{Ben99,daghofer}, it has been also recognized
in the electronic structure calculations performed within the local
density approximation (LDA) that the local interactions are important
and one has to use either LDA+$U$ or LDA+DMFT \cite{yin,Yam06}, i.e.,
take into account also the dynamical aspects in the dynamical mean field
theory. These calculations confirmed the earlier point of view
\cite{Ben99,feiner99} that electron-electron interactions and
the Jahn-Teller terms support each other and both are necessary
to reproduce the observed orbital order in the undoped compound.
In particular, by the studying the effect of pressure it was
concluded that LaMnO$_3$ is not a Mott-Hubbard insulator, but rather
a Jahn-Teller insulator, however, it has been recognized that the 
correlations play also an important role \cite{Yam06}.

The paper is organised as follows. First, we introduce (in
\sref{sec:2}) a {\it realistic\/} model for $e_g$ electrons in the
perovskite manganites which includes the electron Coulomb interactions,
the coupling between $e_g$ electrons with $s=1/2$ spins and
$t_{2g}$ core $S=3/2$ spins, superexchange interaction, and the
interactions of $e_g$ electrons with local distortions due to the
Jahn-Teller terms (\sref{sec:model}). In this section we also present the
method to determine approximate ground states using a $4\times
4\times 4$ cluster. We treat electron correlation effects in the
ground state beyond the Hartree--Fock (HF) approximation using the
so-called local approach (\sref{sec:clust}). The numerical results are 
presented and analysed in \sref{sec:num}. Here we address a question which
values of the parameters of the effective Hamiltonian are
appropriate to account for the experimentally observed situation
and present results obtained for the undoped LaMnO$_3$, and for
La$_{1-x}$Sr$_x$MnO$_3$ at doping levels $x=0.125$, 0.25 and 0.50.
Finally, \sref{sec:summa} contains a short summary of the paper and
general conclusions.

\section{The model and calculation method}
\label{sec:2}

\subsection{The model for $e_g$ electrons}
\label{sec:model}

We study strongly correlated electrons in undoped and doped 3D
perovskite manganites La$_{1-x}$Sr$_x$MnO$_3$
(or La$_{1-x}$Ca$_x$MnO$_3$) using effective model describing only Mn
sites renormalised by surrounding oxygens. At each site we consider
Wannier orbitals of the $e_g$ character composed out of manganese and
oxygen orbitals \cite{yin}. One of them is filled at Mn$^{3+}$ ions,
and both are empty at Mn$^{4+}$ ions in a doped system.
In both cases the $t_{2g}$ orbitals on Mn ions are occupied by three
"core" electrons with total spin $S=3/2$ (treated as frozen and
classical). Thus the active electrons here are only $e_g$ electrons.
We investigate the system by using a model Hamiltonian
\begin{equation}
{\cal H} = H_{\rm kin} + H_{\rm int}+ H_{\rm spin}+ H_{\rm JT} \label{model}
\end{equation}
which consists of the kinetic energy, the on-site Coulomb
interactions, spin interactions, and the Jahn-Teller term.
This model and/or the essential parts of it, were studied earlier
by several groups, see refs.
\cite{dag01,popovic,hotta,daghofer,yin,lin08,jung,rosc03,rosc05,rosc07,rosc08}.

The kinetic part $H_{kin}$ is expressed using a local $e_g$ orbital
basis at each site, which reads in short notation,
\begin{equation}
\label{egbasis}
|z\rangle\equiv (3z^2-r^2)/\sqrt{6}, \hskip .7cm
|x\rangle\equiv ( x^2-y^2)/\sqrt{2}.
\end{equation}
Using this local $e_g$ orbital basis at each site one finds
anisotropic phase-dependent hopping \cite{hotta,daghofer,yin}
\begin{eqnarray}
H_{\rm kin}= -\frac{1}{4} t_0 \sum_{ \{i j\} ||ab, \sigma} \left\{
 (  3 d^{\dagger}_{i x \sigma}  d_{j x \sigma}  +
     d^{\dagger}_{i z \sigma}  d_{j z \sigma} )
\pm \sqrt{3} ( d^{\dagger}_{i x \sigma}  d_{j z \sigma}\right. \\ \nonumber
\left. + d^{\dagger}_{i z\sigma}  d_{j x \sigma} ) \right\}
- t_0 \sum_{ \{i j\} ||c, \sigma} d^{\dagger}_{i z \sigma}  d_{j z \sigma}.
\end{eqnarray}
Here $d^{\dagger}_{i \mu \sigma}$  are creation  operators for an
electron in  orbital $\mu = x,z $
with spin $\sigma=\uparrow,\downarrow$ at site $i$.
The $\{i,j\}$ runs over pairs of nearest neighbours and gives two
contributions for each bond $\langle i,j\rangle$ ; $\pm$ is interpreted
as plus sign for the bond $\langle i,j\rangle$ being parallel to the
crystal axis $a$ and minus for the bond $\langle i,j\rangle$ parallel
to the axis $b$. (The crystallographic axis $c$ is assumed to
be aligned with Cartesian $z$-axis.)

The $H_{\rm int}$ and $H_{\rm spin}$ terms follow from the Coulomb
interactions within a degenerate $d$ band \cite{Ole83}; these
terms are:
\begin{eqnarray}
H_{\rm int} =
  U_0 \sum_{i \mu}  n_{i \mu \uparrow} n_{i \mu \downarrow}
 +
(U_0 - \frac{5}{2} J_H) \sum_{i}  n_{i x} n_{i z}, \\
 H_{\rm spin} =
 - \frac{1}{2} J_H \sum_i (n_{i x \uparrow} - n_{i x \downarrow}
)
(n_{i z \uparrow} - n_{i z \downarrow} )  \\
- J_H \sum_{i \mu} S_i^z  (n_{i \mu \uparrow} - n_{i \mu
\downarrow})  +
\frac{1}{2} J' \sum_{\langle i j \rangle} S_i^z S_j^z, \\
\end{eqnarray}
Here $H_{\rm int}$ describes the charge interactions within the $e_g$
subsystem and plays a role in doped manganites. The spin
interactions $H_{\rm spin}$ include the leading part of Hund's
exchange $J_H$, both between two $e_g$ electrons at the same site,
and between $s=1/2$ spin of a single $e_g$ electron and the
$S=3/2$ core $t_{2g}$ spin at each site. On-site intraorbital
Coulomb interaction is denoted as $U_0$, and the interorbital
Coulomb interaction is a linear combination of this term and
Hund's exchange $J_H$ \cite{Ole83}. Note that we include here only
the leading Ising part of Hund's interaction terms, which is an
approximation. However, within the present method using the
correlated wave functions which are obtained by correcting the
Hartree-Fock (HF) approximation, the remaining transverse terms do
not contribute, compare refs. \cite{rosc07,rosc08}). The term
$\propto J'$ is the antiferromagnetic (AF) superexchange generated
by charge excitations of $t_{2g}$ electrons, which lead to AF
Heisenberg interaction between frozen core $t_{2g}$ electrons
\cite{yin,hotta,feiner99}. This frozen-core approximation for
$S=3/2$ core spins works well for the description of the ground
state properties as shown by earlier studies, see e.g. refs.
\cite{daghofer,weisse}.

Finally, the Jahn-Teller $H_{\rm JT}$ part is
\cite{Bal00,dag01,popovic,lin08,salafranca}:
\begin{eqnarray}
H_{\rm JT} = \sum_i \Big\{
g_{\rm JT}\Big( Q_{1 i}(n_{i x}+n_{i z})+Q_{2 i}\tau_i^x+Q_{3 i}\tau_i^z  \Big) \\
+ \frac{K}{2} \big(2 Q_{1 i}^2  + Q_{2 i}^2+ Q_{3 i}^2 \big)  \Big\}.
\end{eqnarray}
and includes three different Jahn-Teller modes
$\{Q_{1i},Q_{2i},Q_{3i}\}$ at each site. The operators $\tau_i^{\alpha}$
in $H_{\rm JT}$ are constructed for $\tau=1/2$ orbital pseudospin for
$e_g$ electrons at site $i$ \cite{dag01,hotta,daghofer,yin,lin08},
\begin{eqnarray}
\tau_i^x &\equiv &\sum_\sigma
( d^{\dagger}_{i x \sigma}  d_{i z\sigma} +
d^{\dagger}_{i z \sigma}  d_{i x\sigma} ), \nonumber \\
\tau_i^z &\equiv &\sum_\sigma
( d^{\dagger}_{i x \sigma}  d_{i x\sigma} -
d^{\dagger}_{i z \sigma}  d_{i z\sigma} ),
\end{eqnarray}
while $Q_{1 i}$,  $Q_{2 i}$ and $Q_{3 i}$ denote the active Jahn-Teller
deformation modes of the $i-$th octahedron.
(For simplicity the harmonic constant of the isotropic Jahn-Teller
(breathing) mode $Q_1$ is assumed to be double with respect to those
corresponding to $Q_2$ and $Q_3$ unsymmetric modes as discussed in
ref. \cite{dag01}). The breathing mode is quite often neglected in
the effective models for the perovskite manganites, and plays no role
in the undoped system. As we are dealing here with doped systems,
we decided to keep it to investigate its consequences.

The crystal field splitting between $z$ and $x$ orbitals
\eref{egbasis}
was assumed to be zero (this approximation
seems reasonable in quasi-cubic 3D manganites \cite{lin08}
though some authors \cite{yin} assume instead small finite values).
Below we describe the numerical simulatons performed on finite 3D
clusters with the above model \eref{model}
The variants of the described model reproduced the observed sequence
of magnetic phases for increasing hole doping in 2D
monolayer and bilayer manganites \cite{rosc07,rosc08}.

\subsection{Cluster approach including electron correlations}
\label{sec:clust}

We studied $4 \times 4 \times 4$ clusters with periodic boundary
conditions (PBC) filled be different number of $e_g$ electrons,
corresponding to the undoped LaMnO$_3$ system, and to systems with
doping one-eight ($x=0.125$), quarter-doped system ($x=0.25$) and
half-doped system ($x=0.5$). All calculations were performed at
zero temperature ($T = 0$ K). First, the calculations within the
single-determinant HF approximation were performed to determine
the ground state wave function $|\Phi_0\rangle$. Also the energetic
distance $\Delta$ between the highest occupied
molecular orbital (HOMO) and the lowest unoccupied molecular
orbital (LUMO) (HOMO-LUMO gap) was extracted at this step.

In the next step each HF wave functions was independently modified
to improve the energy and to include the electron correlations using
the so-called local ansatz
(for details see refs. \cite{rosc05,rosc03,rosc07,rosc08}). We used
the ansatz for the correlated ground state,
\begin{equation}
|\Psi\rangle = \exp\Big( \sum_i \eta_iO_i\Big)|\Phi_0\rangle\,,
\label{la}
\end{equation}
where $\{\eta_i\}$ are variational parameters, and $\{O_i\}$ is a
set of local correlation operators. These operators include the
leading local density-density correlations in the present model
for the perovskite manganites, in analogy to the 2D layered
manganites, see refs. \cite{rosc07,rosc08}. These local operators
used in the present model correspond to the subselection of the
most important two electron excitations within the {\it ab
initio\/} configuration--interaction method.  The variational
parameters $\{\eta_i\}$ were found by minimising the total energy
\begin{equation}
E_{\rm tot}=\frac{\langle\Psi|H|\Psi\rangle}{\langle\Psi|\Psi\rangle}\,.
\label{etot}
\end{equation}
In this way the correlation energy,
\begin{equation}
E_{\rm corr}=E_{\rm tot}-E_{\rm HF}, \label{ecorr}
\end{equation}
was obtained.

Coming to technical details, in the first step HF computations
were performed for each considered electron filling, starting from
one of several different initial conditions (about a thousand for
each set of Hamiltonian parameters), i.e., from predefined (some
symmetric but mostly random) charge distribution, spin
configuration, the latter selected from eight predefined patterns
of core $t_{2g}$ spins (see figure 1 of ref. \cite{rosc08}), and
from several predefined sets of classical variables
$\{Q_{1i},Q_{2i},Q_{3i}\}$. For each fixed set of the starting
parameters and starting initial conditions on convergence we
obtained a HF wave function which was a candidate for the ground
state wave function. This self-consistent procedure was performed
to provide energy minimum also with respect to the
$\{Q_{1i},Q_{2i},Q_{3i}\}$ classical variables
\cite{rosc07,rosc08}.

After completing the HF computations we performed correlation
computations (as a second step for each investigated wave
function) obtaining the total energy \eref{etot} and the
correlation energy \eref{ecorr} (for more details see refs.
\cite{rosc05,rosc03,rosc07}). After obtaining the total energy for
one configuration, we repeated the entire procedure from the
beginning, i.e., we took another (second) set of HF initial
conditions and repeat all computations to obtain the second
candidate for the ground state. Similarly for the third, fourth,
$\cdots$, {\it etcetera\/}, set of initial conditions. Finally,
the resulting set of total energies was inspected and the lowest
one was identified as a good candidate for the true ground state.

\section{Numerical results}
\label{sec:num}

\begin{figure}[t!]
\begin{center}
\includegraphics[width=8cm,angle=-90]{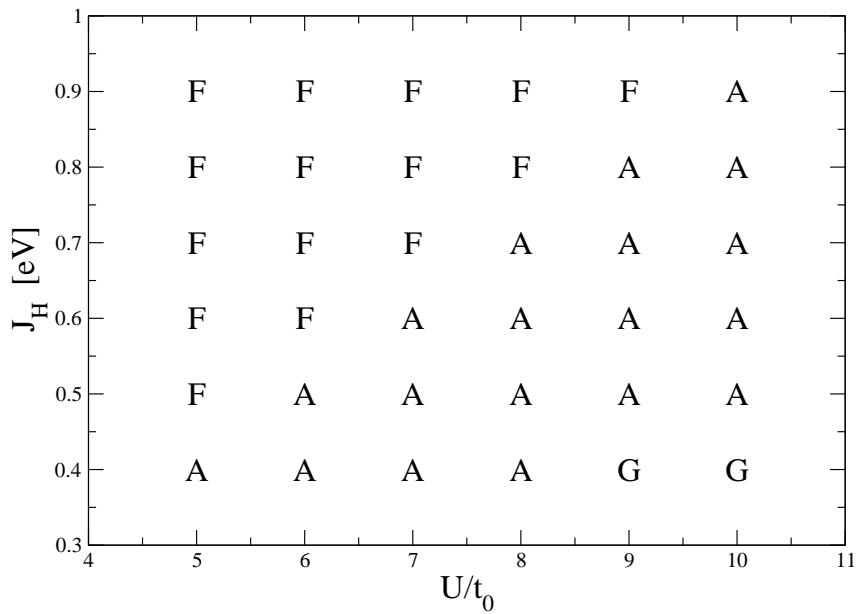}
\end{center}
\caption{ A simplified phase diagram in $(U,J_H)$ plane for
undoped ($x=0$) LaMnO$_3$. $G$ refers to an ordinary N\'eel
antiferromagnet ($G$-AF phase), $A$ to $A$-AF phase with FM planes
and AF coupling along the direction perpendicular to them, and $F$
stands for the FM phase. Parameters: $t_0=0.4$ eV, $g_{\rm
JT}=2.2$ eV \AA$^{-1}$, $K=13$ eV \AA$^{-2}$ and $J'=3$ meV. }
\end{figure}

As the parameters of the model \eref{model} are known only with
certain accuracy, the computations were repeated for many sets of
the Hamiltonian parameters. The values of the parameters which one
should consider as realistic for La$_{1-x}$(Sr,Ca)$_x$MnO$_3$ were
extensively discussed in the literature (for a review in context
of the present paper see refs.
\cite{dag01,daghofer,rosc07,rosc08}). Following the standard
parameter sets \cite{rosc07,rosc08}, we assumed that $t_0=0.4$ eV
and the intraorbital Coulomb element is limited in the range $ 5 <
U/t_0 < 12$. The values of Hund's exchange are larger than $t_0$
\cite{kovaleva,feiner99,Ole02}, and we considered $ t_0 < J_H < 2
t_0$. The Jahn-Teller constant $K$ was fixed as $K = 13$ eV
\AA$^{-2}$ \cite{dag01,millis,bala2002,rosc03,rosc07}. Not much is
known about the value of the coupling constant $g_{\rm JT}$,
therefore following discussion in ref. \cite{dag01} we assumed
that 2 eV \AA${}^{-1}$ $ < g_{\rm JT} <$ 3.7 eV \AA${}^{-1}$. And,
finally, following recommendation of refs. \cite{lin08,feiner99}
we use the experimental value of the N\'eel temperature in
CaMnO${}_3$ \cite{3meV}, where $e_g$ electrons are absent and the
isotropic superexchange between core $t_{2g}$ spins follows solely
from the excitations of $t_{2g}$ electrons, to fix $J'=3$ meV.
Thus we have three fixed and three free parameters.

\begin{figure}[t!]
\begin{center}
\includegraphics[width=8.0cm,angle=0]{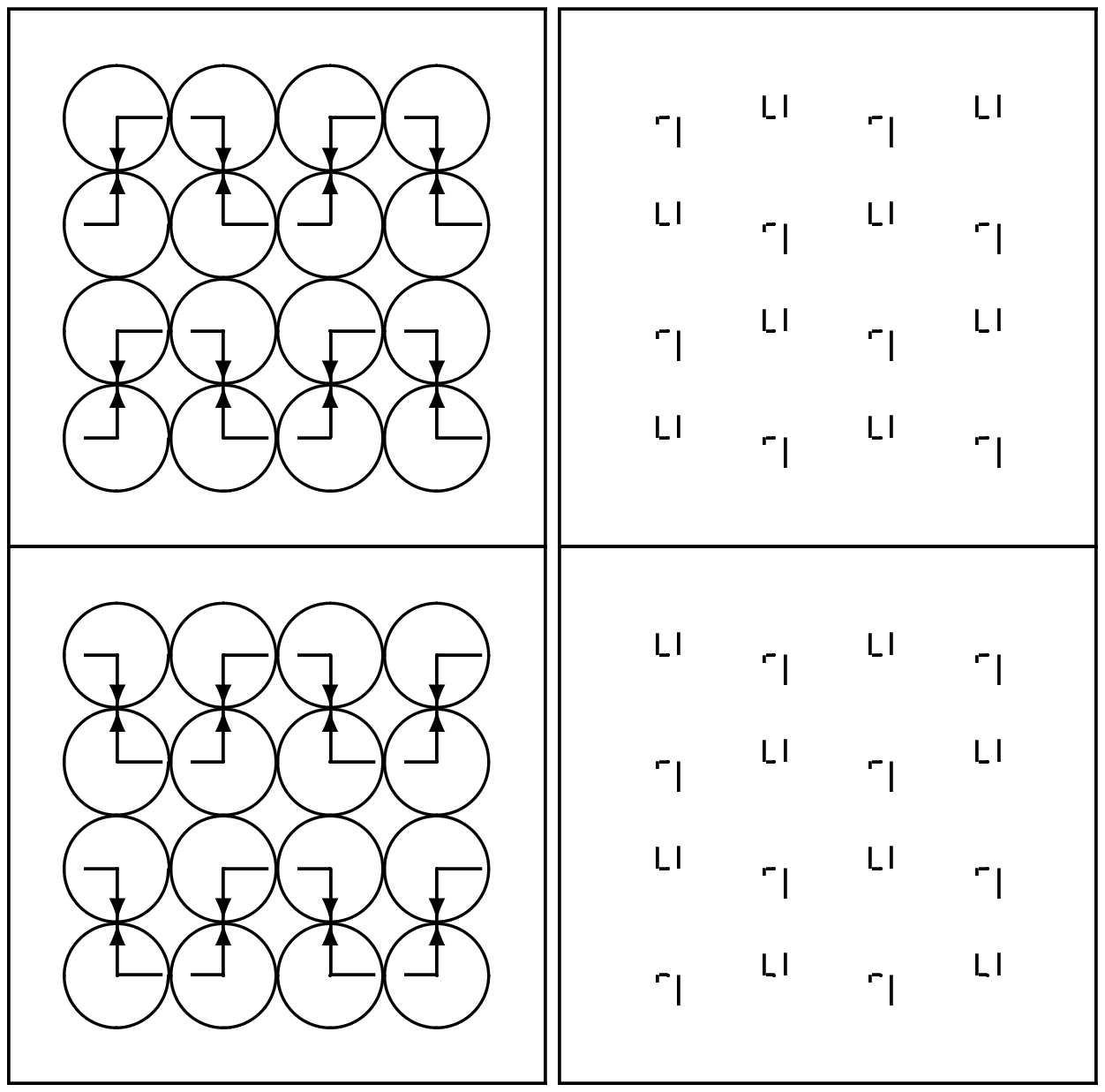}
\end{center}
\caption{%
Charge distribution, magnetic, and orbital order of $e_g$
electrons in the $A$-AF phase of LaMnO$_3$, as obtained in undoped
$4 \times 4 \times 4 $ cluster with PBC. charge and alternating
$x$, $z$ orbital order for $e_g$ electrons Only single $ab$ planes
are shown. Core $t_{2g}$ spins are not shown.
Left side panels --- At each site the circle diameter corresponds
to total $e_g$ on-site charge; the arrow length to the $e_g$ spin;
the horizontal bar length to charge density difference between $x$
and $z$ orbitals (longest bar to the right corresponds to pure
$x$, to the left to pure $z$, zero length to half by half
combination). All these values are expressed in approximate
proportionality (as generated by graphic package of latex) to
nearest neighbour site-site distance which is assumed to be unity.
Upper panels show situation on even numbered $ab$ planes, lower
panels correspond to odd numbered $ab$ planes.
Right side panels --- Jahn-Teller distortions for the same even and odd
$ab$ planes: $Q_{2i}$ (bars drown slightly to the left of each
site) and $Q_{3i}$ (bars to the right of each site), and isotropic
$Q_1$ (breathing) mode in between. For more clarity $Q_2$ and
$Q_3$ bars are enlarged by factor of 2 compared to $Q_1$. In the
present case all $Q_{1i}$ take the value of -0.08 \AA, the largest
$Q_{2i}$ is equal to 0.10 \AA, the largest $Q_{3i}$ is 0.14 \AA. }
\end{figure}

The aim was to identify the set (or sets) of these three free
parameters which lead to good agreement of the model predictions
with experimental data. This task turned out to be rather
difficult especially with respect to Jahn-Teller parameter $g_{\rm
JT}$. Actually almost all values of $g_{\rm JT}$ either yield
conducting state for both $x=0.125$ and $x=0.25$ doping levels,
or they yield an insulator, also for both above doping levels. This
contradicts the experimental situation as
La$_{1-x}$Sr$_x$MnO$_3$ manganite is weakly insulating for $x=0.125$
and conducting for $x=0.250$ \cite{resistance}. This
insulator-to-metal transition turned out to be very useful as it
allowed us to fix all three free parameters. The values of the
parameters which can bring out the agreement with the experimental
data form a small isolated region within the parameter space, and
finding them required making hundreds of scans through the
parameter space. We have found that only for $U/t_0=8$,  $J_H=0.7$
eV and $g_{\rm JT} = 2.2$ eV \AA${}^{-1}$ we are able to reproduce
the experimental results.

\subsection{Undoped LaMnO$_3$}
\label{sec:0}

The insulating phase in LaMnO$_3$ was reproduced by the present
calculations. For the magnetic order we obtained rubust $A$-AF
structure, which has ferromagnetic (FM) order in $ab$ planes,
coupled by AF interaction in the $c$ direction. This phase is
almost generic and can be reproduced on the phase diagram for many
sets of Hamiltonian parameters. Morover, the correlation energy is
small (see Table 1) and does not change the magnetic and orbital order
obtained from the HF wave functions. To give an example, let us show
one 2D section of our simplified phase diagram. Note that as long
as we ignore the global structural distortion of the lattice below
that structural phase transition, all cubic directions are
equivalent and $A$-AF order shown in figure 1 applies also e.g. to
FM order in $ac$ planes accompanied by the AF coupling along the
$b$ direction.

As already mentioned the experimental data on conductivity fix the
electron interaction parameters $U/t_0 = 8$ and $J_H = 0.7$ eV, as
well as $g_{\rm JT} = 2.2$ eV \AA${}^{-1}$. For these parameters
the ground state ordering for LaMnO$_3$ is shown in figure 2.
The Jahn-Teller $Q_2$ and $Q_3$ distortions measured in ref.
\cite{chatterji} (at $T = 200$ K) are only roughly in agreement
with our results. (Note that for simplicity the cubic Jahn-Teller
terms \cite{popovic,lin08} in $H_{\rm JT}$ were omitted.
Therefore, the Jahn-Teller distortions we obtained are only estimations
rather than the precise values).

\begin{figure}[t!]
\begin{center}
\includegraphics[width=8cm,angle=0]{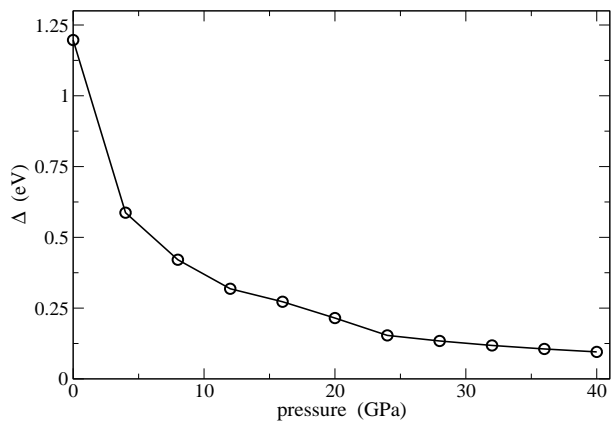}
\end{center}
\caption{
HOMO-LUMO gap $\Delta$ for increasing pressure $p$ in the undoped
LaMnO$_3$: results of the calculations are shown by circles, solid
line is a guide to the eye. The linear dependence of $t_0(p)$ and
$K(p)$ on pressure is assumed. Other Hamiltonian parameters take
standard values, see the caption of figure 2. }
\end{figure}

We have found a large HOMO-LUMO gap $\Delta$ for LaMnO$_3$ which is
an indicator that the obtained $A$-AF phase is insulating, see figure
3. It is known that at pressure of 32 GPa LaMnO$_3$ becomes
conducting \cite{loa}. In order to investigate the effect of
pressure we followed heuristic argumentation presented in ref.
\cite{32gpa} --- it has been argued there (see figure 2 in ref.
\cite{32gpa}) that the effect of 32 GPa pressure can be reproduced
in a model similar to our effective model \eref{model} when $t_0$
and $K$ Hamiltonian parameters are renormalised in the following
way: $t_0\rightarrow 1.4t_0$ and $K\rightarrow 3.3K$. Following
ref. \cite{32gpa}, the other parameters (of the model) were left
unchanged. We note that this assumption is somewhat unrealistic as
at least the superexchange $J'$ would also increase under pressure,
but this effect is not important as long as we investigate the ground
state and the magnetic structure does not change. $U_0$ and other
effective parameters of the Hamiltonian should also show some
limited variation under increasing pressure.

Inserting the new renormalised values of hopping and
Jahn-Teller constant into our computations we found that the high
pressure phase is indeed a conductor and we could identify the
conducting phase as perfect charge-homogeneous, FM phase with zero
Jahn-Teller distortions and with perfect half-to-half $(x^2-y^2)$
and $(3z^2-r^2)$ orbital order on each individual site. This
orbital liquid state occurs in the doped FM manganites under
normal pressure \cite{Fei05}.

To provide a better evidence to such a prediction we plotted
HOMO-LUMO gap $\Delta$ for increasing pressure. For simplicity it was
assumed that $t_0(p)$ and $K(p)$, where $p$ denotes pressure, vary linearly
from $t_0 $ and $K$ at $p=0$ to $1.4 t_0$ and $3.3K$ at $p=32$ GPa. Anyway,
the transition from insulating to conducting phase can be deduced from
figure 3 and it occurs around $p\approx 30$ GPa where the gap
becomes small enough, $\Delta\simeq 0.1$ eV. Note that in the present
finite cluster one finds always a gap between HOMO and LUMO, but we
have estimated that $\Delta\simeq 0.1$ eV corresponds to a metallic
ground state in the bulk.
To close the discussion on LaMnO$_3$ under pressure
let us mention ref. \cite{nan10} where
magnetic and orbital order under uniaxial strain
(only JT modes were pressure-renormalised) was studied
by using similar (to our) Hamiltonian.

\begin{figure}[t!]
\begin{center}
\includegraphics[width=8.4cm,angle=0]{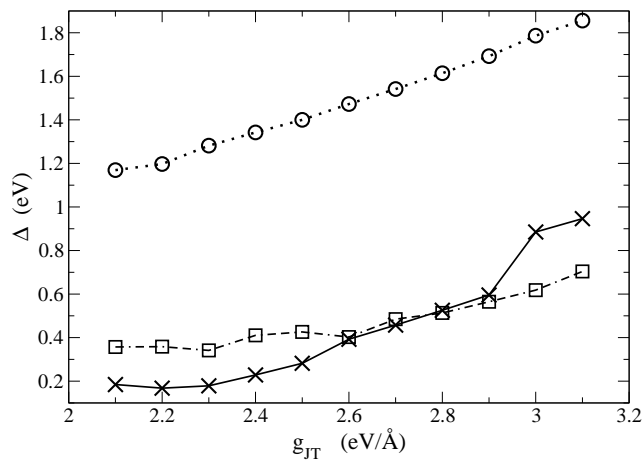}
\end{center}
\caption{ HOMO-LUMO gap $\Delta$ for increasing Jahn-Teller coupling
$g_{\rm JT}$. Circles, squares and crosses correspond to $x=0$,
$x=0.125$ and $x=0.25$, respectively. Parameters: $t_0 =0.4$ eV,
$U/t_0 = 8$, $J_H =0.7$ eV, $J'=3$ meV.}
\end{figure}

\subsection{One-eight-doped La$_{0.875}$Sr$_{0.125}$MnO$_3$}
\label{sec:1/8}

In one-eight doped manganite one finds an insulating phase with FM
order, both in the HF and for the correlated wave functions. The
correlation energy is small, see Table 1, and does not change the
magnetic and charge order in the system.
The breathing mode Jahn-Teller distortions are small and uniform,
$Q_{1i} \approx -0.08$ \AA. This phase is characterized by a rather
interesting charge and orbital distribution, with a distinct alternation
in charge density between even and odd planes. Consequently, this phase
is an insulator, in agreement with the experiment \cite{resistance},
though a weak one as indicated by rather small HOMO-LUMO gap shown in
figure 4 (curve in the middle, squares). Note that gap is reduced by a
factor close to 3 from the value obtained for LaMnO$_3$.

\Table{\label{tab:ene} Hartree-Fock energies $E_{\rm tot}$ and
total ground state energies $E_{\rm HF}$ \eref{etot}, both in eV,
as obtained for 4 $\times$ 4 $\times$ 4 cluster versus the doping
$x$. Parameters of the Hamiltonian: $t_0= 0.4$ eV, $U =8 t_0 $,
$J_H = 0.7$ eV, $K=13$ eV/\AA${}^2$, $g=2.2$ eV/\AA, $J'=3.0$
meV.}
%
%
%
\begin{tabular}{ccc}
  $x$   & $E_{\rm HF}$ & $E_{\rm tot}$ \\  \hline
0.000   &  -96.016     &  -96.114  \\
0.125   &  -88.217     &  -88.315  \\
0.250   &  -79.374     &  -79.521  \\
0.500   &  -58.666     &  -58.896  \\  \hline
\end{tabular}
%
\endtab

The alternation of $e_g$ electron density occurs between even and
odd planes. First, in even numbered planes there are somewhat less
electrons and the orbitals are of $(x^2-y^2)$ character,
$Q_{2i}\approx 0$ and $Q_{3i}$ take the value of $-0.1$ \AA. These
distortions favor the $x$ orbitals. However, there are also some
exceptions, i.e., occasionally on a site we find the $(3z^2-r^2)$
orbital occupied and $Q_{3i}\approx +0.1$ \AA. The charges are
distributed in a slightly non-homogeneous way, with the site to
site variations of charge being at most 0.2 $e$.
Second, in the odd numbered planes one finds a little bigger
average charge density per site than in the even planes and almost
homogeneous charge distribution. In contrast to the even planes,
one finds predominantly $(3z^2-r^2)$ orbitals occupied. The
Jahn-Teller distortions are $Q_{3i} \approx \pm 0.1$ \AA \; and
$Q_{2i} \approx \pm 0.1$ \AA. Again, we have seen also some
exceptions, with $x$ orbitals occupied at a few sites and the
corresponding modified Jahn-Teller distortions, similar to those
in the even planes. In fact, this result shows that sites with
differently occupied orbitals can proliferate between the even and
odd planes and form charge and orbital defects in each plane. This
shows that the inhomogeneous charge distribution is a generic
feature in this range of doping.

\subsection{Quarter-doped La$_{0.75}$Sr$_{0.25}$MnO$_3$}
\label{sec:1/4}

\begin{figure}[t!]
\begin{center}
\includegraphics[width=8.0cm]{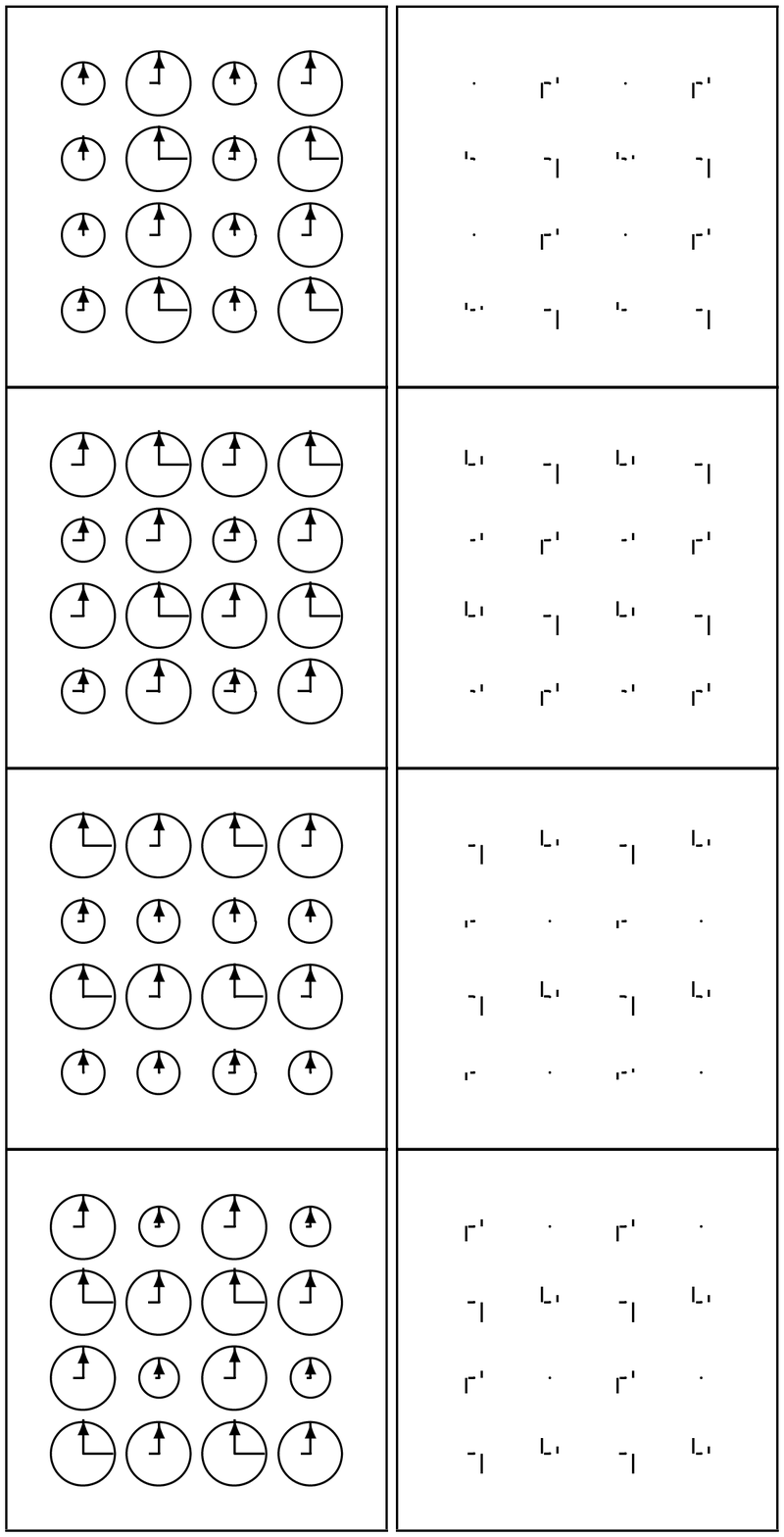}
\end{center}
\caption{%
Magnetic A-AF, charge and orbital electron density for $e_g$ electrons as
obtained in quarter-doped manganites, $x=0.25$. Four single $ab$
planes (perpendicular to the $c$ axis) extracted from $4 \times 4
\times 4$ cluster are shown. Core $t_{2g}$ spins are not shown;
their directions agree with the directions of $e_g$ spins. The
meaning od symbols and parameters are the same as in figure 2.
}
\end{figure}

Also for quarter-doped manganites, such as
La$_{0.75}$Sr$_{0.25}$MnO$_3$, one finds FM order and weakly
non-homogeneous charge distribution with the present parameters
(see figure 5). These results are similar to the ground state
found for $x=0.125$ --- we have found again nonequivalent
alternating planes, as described below. The correlations do not change
this result, and are again weak (Table 1) as the local moments
have formed and the main effects are already captured in the HF
approach.

On the one hand, within even numbered planes in the ground state
one has weak charge order in form of a stripe phase, with vertical
or horizontal charge-minority lines alternating with
charge-majority lines (charge difference of about $0.2e$). Charge
minority sites have mixed half to half $x$ and $z$ orbital
occupancy and negligible Jahn-Teller distortions, which is
characteristic for the orbital liquid in the metallic state
\cite{Fei05}. The situation is quite different along the charge
majority lines --- one finds there half of these sites with $x$
orbitals and with negligible $\{Q_{1i},Q_{2i}\}$ distortions, and
with $Q_{3i}\approx -0.13$ \AA. The other half of these sites
carries the mixture of $z$ and $x$ roughly in proportion 2:1 and
negligible $\{Q_{1i},Q_{3i}\}$ distortions, whereas $Q_{2i}\approx
-0.1$ \AA. These sites with higher electron density (i.e., those
with $x$ orbitals and those with 2:1 orbitals $z$ and $x$)
alternate along the charge-majority lines.

On the other hand, on odd numbered planes the orbitals and
Jahn-Teller distortions are much the same as the ones on even
numbered planes (as described above). However there is a
difference --- along charge-minority lines (each second line in
the cluster) we did not find homogeneous charge density but,
instead alternating charge-minority and charge-majority sites.

The FM phase at $x=0.25$ doping should be a metal which follows
from quite small HOMO-LUMO gap  ($\Delta\simeq 0.1$ eV), see
figure 4. This result agrees with the phase diagram for the
perovskite manganites \cite{dag01,Che95}.

\subsection{Half-doped La$_{0.5}$Sr$_{0.5}$MnO$_3$
            --- stability of the CE phase}
\label{sec:1/2}

Surprisingly, for half-doped La$_{0.5}$Sr$_{0.5}$MnO$_3$ one finds
again almost the same ground state as for $x=0.25$, namely FM spin
order, accompanied by weakly non-homogeneous charge distribution
and a conducting phase (HOMO-LUMO gap is about 0.12 eV). The ratio
of $x$ to $z$ orbitals on each site is half to half, while the
Jahn-Teller distortions are very small. These findings reproduce
the experimental situation in La$_{0.5}$Sr$_{0.5}$MnO$_3$, where
no CE phase was observed.

\begin{figure}[t!]
\begin{center}
\includegraphics[width=8.0cm]{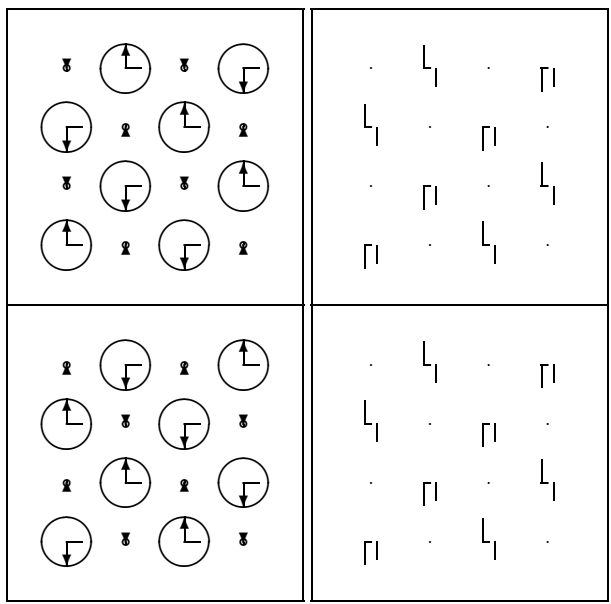}
\end{center}
\caption{%
Magnetic CE, the ordered charges and fractions of occupied $x$ and
$z$ orbitals for $e_g$ electrons in half-doped
La$_{0.5}$Ca$_{0.5}$MnO$_3$. The standard Hamiltonian parameters
(which apply to strontium-doped substance) were assumed as in the
caption of figure 2 with one exception, namely we adopted a larger
Jahn-Teller coupling $g_{\rm JT} =3.7$ eV \AA${}^{-1}$. Two panels
correspond to two neighboring nonequivalent $ab$ planes
--- the order found in these planes is repeated along the $c$ axis
in the planes number 3 and 4, not shown.
Only $e_g$ spins are shown ($T_{2g}$ spins are parallel to them at
all sites). Note that the magnetic CE zig-zag-like order refers to
the total spin thus it is not properly visible on the figure where
only $e_g$ spins are plotted. The legend is the same as in figure
2. }
\end{figure}

On the contrary, the CE phase was found in half-doped
La$_{0.5}$Ca$_{0.5}$MnO$_3$ \cite{ce1,ce2}. A somewhat unprecise
but still common point of view is that a manganite doped with
strontium is more metallic than the manganite doped with calcium
(more insulating) which in turn might be attributed to weaker
coupling to the lattice by the Jahn-Teller interactions in
strontium compounds and a bigger role of these interactions in
calcium compounds. Indeed, the FM phase was found insulating in a
broader range of doping in La$_{1-x}$Ca$_x$MnO$_3$ than in
La$_{1-x}$Sr$_x$MnO$_3$ \cite{Mou07}. If this is the case then our
effective model should provide a different prediction for the
ground state when $g_{\rm JT}$ is larger.

When we discussed the precise values of Hamiltonian parameters
following the discussion presented in ref. \cite{dag01} we assumed
2 eV \AA${}^{-1}$  $ < g_{\rm JT} <$ 3.7 eV \AA${}^{-1}$.  Then
for strontium manganite La$_{1-x}$Sr$_{x}$MnO$_3$ we found that
$g_{\rm JT} =  2.2$ eV \AA${}^{-1}$. Now for the sake of the ongoing
discussion for calcium doped manganite we will accept $g_{\rm JT}=3.7$
eV \AA${}^{-1}$, the biggest value which is allowed. This change
of the unique parameter immediately provides us with the expected
result. Now the spin order in the ground state is indeed in form
of FM zig-zags with AF order between them, characteristic for the CE
phase, and the charge order is of checkerboard type (figure 6).
The CE phase is quite robust, it is identified at the HF level and
the correlations when added do not change this result. This is in
contrast to what happens in bilayer half-doped manganites where
the correlations stabilize the CE phase \cite{rosc08}. We note,
however, that the magnetic moments are larger in the 3D system,
and therefore the role of electron correlations in the magnetic
states such as the CE phase is reduced.

\subsection{Switching off Jahn-Teller distortions}
\label{sec:JT}

The results presented up to now indicate the importance of
Jahn-Teller interactions. Due to them local distortions are
created at each doping level, and they influence both the
electronic structure and the magnetic properties. It seems that
the value of $g_{\rm JT}$ coupling is the primary factor which
determines whether the 3D perovskite system is a conductor or an
insulator. Therefore it seems reasonable to perform additional
test computations switching off the Jahn-Teller  interactions.

We performed such computation for La$_{1-x}$Sr$_{x}$MnO$_3$ where
$x= 0, 0.125, 0.250$ and  $g_{\rm JT} =0$. The resulting ground
states were all the same: FM, charge homogeneous
orbital liquid phase \cite{Fei05}, with half to
half ratio of $x$ to $z$ orbitals at each site and conducting
(HOMO-LUMO gaps we found were all about 0.1 eV). This confirms the
above point of view that the Jahn-Teller distortions play a crucial
role for the charge, magnetic and orbital order in doped perovskite
manganites.

\section{General conclusions and final remarks}
\label{sec:summa}

We considered the effective Hamiltonian for $e_g$ electrons which
takes into account their kinetic energy with anisotropic phase
dependent hopping $\propto t$, on-site Coulomb interaction $U$,
Hund's exchange coupling $J_H$, their coupling to frozen core
$t_{2g}$ electrons $\propto J_H$, the Heisenberg AF superexchange
$\propto J'$ between core $T_{2g}$ spins, and the Jahn-Teller
interactions with lattice distortions.
Due to the complexity of the problem, we could not provide a
precise proof that this model is sufficient for the physical
properties of the perovskite manganites, but rather we report on
results of computations which demonstrate that all these
interactions enter on equal footing and seem to be essential for
correct interpretation of the experimental data. Indeed, we have
performed also calculations with several sets ignoring one or the
other interaction term, and the results were very unsatisfactory
in all cases, i.e., such simplified models were not able to
reproduce the experimental observations.

We would like to emphasize that the realistic Jahn-Teller coupling
plays a rather special and a very important role in the manganites
as it controls the conductivity of the perovskite 3D systems. We
have presented evidence that this interaction generates local
distortions in the entire investigated range of hole doping
$0.125\le x\le 0.50$ which lead to charge inhomogeneities, defects
in the orbital occupancy, and influences the magnetic order. Such
states have been seen before in the experimental data on doped
manganites \cite{Lou99} and were also suggested to follow from the
theoretical models \cite{Ali01}. Unfortunately, reliable studies of
systems with charge disorder and local distortions are very difficult
to perform, but the present results demonstrate that these states
might decide about the colossal magnetoresistance and select one or
the other type of magnetic order. Particularly at half-doping ($x=0.5$)
we have found a subtle balance between the FM phase and the CE phase,
the latter stabilized by large Jahn-Teller interactions.
This is not surprising as the nearest neighbor JT coupling might not 
be sufficient to stabilize the observed zigzag FM chains in the CE 
phase for the realistic parameters. The mechanism invoked in \cite{Hor04} 
to stabilize the CE phase is subtle and employes the cooperative JT 
interaction between next-nearest Mn$^{3+}$ neighbors mediated by the 
breathing mode distortion of Mn$^{4+}$ octahedra and displacements 
of Mn$^{4+}$ ions.

As found before for monolayer and bilayer doped manganites, see
refs. \cite{rosc03,rosc07}, we have obtained many metastable
ground states in the 3D system which are possible in principle.
Such states are not discussed in detail in the present paper, but
we just point out that they represent local energy minima and
could be also realized at finite temperature. Therefore, the
search for the best candidate for the true ground state requires
examination of numerous phases with different kinds of
nonhomogeneous charge distribution. Therefore, we suggest as a
general conclusion that nonhomogeneous charge distribution is a
generic feature of the doped perovskite manganites. It follows
that the theoretical predictions which are based on models assuming
just a few {\it a priori\/} selected candidates for the ground state
(especially the ones with high symmetry of charge or orbital order)
can not be fully conclusive unless they are based on reliable
experimental data.

\ack

We acknowledge financial support by the Polish Ministry of Science
and Higher Education under Project No. N202~068~32/1481.
A M Ole\'s was also supported by the Foundation for Polish Science (FNP).

\end{document}